\documentclass[showpacs,twocolumn,aps,floatfix,superscriptaddress]{revtex4}
\usepackage{amsmath,amssymb,eucal,graphicx,subfigure,enumerate}
\begin{document}

\def\av#1{\langle#1\rangle}
\def\p{\bullet}
\def\q{\circ}
\def\prob#1{{\rm Pr}(#1)}
\newcommand{\qq}
    {\setlength{\fboxsep}{-.4pt}\fbox{\rule{1.3em}{0mm}\rule{0mm}{1ex}}}
\newcommand{\pp}{\rule{1.3em}{1ex}}

\title{Exact solution of the Nonconsensus Opinion Model on the line}

\author{Daniel ben-Avraham}
\email{benavraham@clarkson.edu}
\affiliation{Physics Department, Clarkson University, Potsdam, NY
13699-5820, USA} 
\affiliation{Department of Mathematics \& Computer Science, Clarkson University, Potsdam, NY 13699-5815}

\begin{abstract}
The nonconsensus opinion model (NCO)  introduced recently by Shao et al., [{\it Phys.~Rev.~Lett.} {\bf 103}, 018701 (2009)]
is solved exactly on the line.  Although, as expected, the model exhibits no phase transition
in one dimension, its study is interesting because of the possible connection with invasion percolation with trapping.
The system evolves exponentially fast to the steady-state, rapidly developing long-range correlations: The average cluster size in the steady state scales as the {\em square} of the initial cluster size, of the (uncorrelated) initial state.  We also discuss briefly the NCO model on Bethe lattices,  arguing that its phase transition diagram is different than that of regular percolation.
\end{abstract}

\maketitle

\section{Introduction}
A Nonconsensus Opinion model (NCO) has been introduced recently~\cite{shao} where each node in a graph can be in one of two states (representing opinions).  At each time step each node adopts the state agreeing with the majority of the
nodes in its {\em neighborhood}, consisting of the node {\em itself} and its nearest neighbors.  In the event of a tie, the node
retains its original state~\cite{remark1}.  The NCO model is thus similar to the majority-voter model, but where self-opinion counts.  This difference is sufficient to ensure survival of the minority opinion, in the steady state~\cite{shao}.  

In ~\cite{shao} a conjecture was made that the NCO model and invasion percolation with trapping (TIP)~\cite{tip} are in the 
same universality class, and this was  well supported by convincing numerical evidence.  As opposed to invasion percolation {\em without} trapping, which is known to be in the same universality class as regular percolation, TIP is much less well understood~\cite{tip}.  The following exact analysis of the NCO model, even though limited to the simplest case of one dimension, is therefore interesting not only on its own right, but also because of the possible connection with TIP.  In addition, we discuss the phase transition diagram of the NCO model on Bethe lattices, finding a rich three-phase diagram, reminiscent of regular percolation in nonamenable {\em one-ended} graphs.  Models similar to the NCO have been considered in the context of social studies~\cite{latane}.

\section{The NCO model in 1D}
In the following we will specialize to the evolution of the NCO model on the infinite line.  Each lattice site
can be either empty ($\q$) or occupied ($\p$).  The initial state is completely random, with a fraction $p_0=p$
of occupied sites, and $q_0=1-p\equiv q$ empty sites.  At each time step all of the sites get updated {\em simultaneously}
according to the NCO rule: the new state of each site agrees with the majority of the states of the site's neighborhood,
consisting of the site itself and its {\em two} nearest neighbors.  Thus, an occupied site remains occupied if at least one
of its neighbors is occupied, but becomes empty in the next time step if both its neighbors are empty.  An empty site remains empty if at least one neighbor is empty, and becomes occupied if both neighbors are occupied.  It is easy then to predict
the density of occupied and empty sites after the first step:
\begin{equation}
\label{p1}
p_1=p(1-q^2)+qp^2=3p^2-2p^3\,,\quad q_1=3q^2-2q^3\,.
\end{equation}
After this first step, the initially random state of the system develops correlations
and one can no longer apply the same reasoning to predict the state of the system in subsequent steps.
Indeed, ignoring correlations and iterating the mapping~(\ref{p1}), one arrives at the {\em erroneous} conclusion that
the final state of the system consists of either all empty or occupied sites (depending on whether $p$ is smaller or larger than $1/2$).

It is clear from the evolution rules that a cluster of two or more occupied (or empty) sites
remains stable forever.  We denote occupied clusters $\p\p\cdots\p$ with a solid rectangular box \pp\,,
and empty clusters $\q\q\cdots\q$ with an empty box \qq\,.  

The initial configuration can be viewed as a collection of stable blocks, separated by alternating sequences of occupied-empty sites.  There are only four possibilities:
\begin{enumerate}[(a)]
\item $\qq\p\q\p\q\cdots\p\q\,\pp$~~($2n$ intervening sites)
\item $\pp\q\p\q\p\cdots\q\p\,\qq$~~($2n$ intervening sites)
\item $\qq\p\q\p\q\cdots\q\p\,\qq$~~($2n+1$ intervening sites)
\item $\pp\q\p\q\p\cdots\p\q\,\pp$~~($2n+1$ intervening sites)
\end{enumerate}
In all of these cases all of the sites in the alternating sequences flip state at each time step.  On each step, the two sites 
at the edges merge with the bounding blocks.  Thus, cases (a) and (b) cease to evolve after $n$ steps, and the bounding
blocks meet, having each grown by $n$ sites.  For example,  $\qq\p\q\p\q\,\pp\to\qq\p\q\,\pp\to\qq\,\pp\,$.  Cases (c) and (d) evolve for $n+1$ steps, after which the bounding blocks merge, having subsumed all of the intervening sites, e.g.,
$\qq\p\q\p\qq\to\qq\p\qq\to\qq\,$. 

\section{The steady state}
At the steady state the system crystalizes into alternating blocks, $\cdots\qq\,\pp\,\qq\,\pp\cdots$, each consisting of at least two sites.  We now compute several characteristics of the steady state.

%
Let us first look at the density of {\em kinks} --- the boundaries between clusters.  We consider kinks of type $\qq\,\pp$ (kinks
of type $\pp\,\qq$ have the same density, by symmetry).

The location of the kinks in the steady state can be predicted completely from the random initial state, as only case (a), above, develops into the type of kink in question.  Thus the steady-state density of kinks, $\kappa_s$, is:
\begin{eqnarray}
\label{ks}
&&\kappa_s=\nonumber\\
&&\prob{\qq\,\pp}+\prob{\qq\!\p\!\q\pp}+\prob{\qq\!\p\!\q\!\p\!\q\pp}+\cdots\nonumber\\
&&=p^2q^2(1+pq+p^2q^2+\cdots)=\frac{p^2q^2}{1-pq}\,.
\end{eqnarray}
Notice that just two adjacent occupied (empty) sites are sufficient to guarantee the existence of a \pp\ (\qq) cluster,
accounting for the overall $p^2q^2$ factor.  That $\kappa_s$ is smaller than $pq$, the density of $\q\p$-kinks in the initial state, reflects the fact that the system coarsens over time.

\begin{figure}[t]
\includegraphics[width=0.45\textwidth]{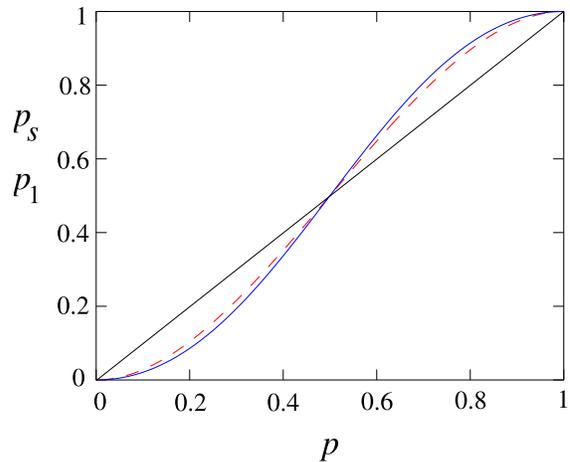}
\caption{(Color online) Density of occupied sites:
Shown are  the density of occupied sites at the steady state, $p_s$, (solid curve) and after one time step, $p_1$, (broken line),
as a function of the initial density, $p$.
The initial density $p$ is plotted as well (straight line), for the sake of comparison.}
\label{ps.fig}
\end{figure}

%
To obtain the steady-state density of occupied sites, $p_s$, we observe that cases~(a) and~(b) keep the number of occupied
and empty sites constant over time.  Case~(d) converts $n+1$ empty sites into occupied sites, while case~(c) does the opposite, emptying $n+1$ occupied sites.
Putting all this together, we get
\begin{eqnarray}
\label{ps}
p_s&=&p+p^4q(1+2pq+3p^2q^2+\cdots)\nonumber\\
&&-q^4p(1+2pq+3p^2q^2+\cdots)\nonumber\\
&=&p+\frac{pq(p^3-q^3)}{(1-pq)^2}\,.
\end{eqnarray}
Thus, unless $p=1/2$, the majority species increases on expense of the minority species (Fig.~\ref{ps.fig}).

%
We now compute the probability distribution for clusters of $n$ occupied sites in the steady state.  
In the initial state, the probability of an $n$-cluster is $P(n)=q^2p^n$, so the expected length is
$\av{n_\p}_0=\sum_{n=1}^{\infty} nP(n)/\sum_{n=1}^{\infty} P(n)=1/q$.

\begin{figure}[t]
\includegraphics[width=0.4\textwidth]{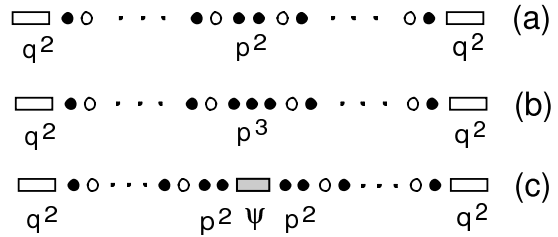}
\caption{Cases contributing to the eventual formation of $n$-clusters of occupied sites (see text).}
\label{clusters.fig}
\end{figure}

Fig.~\ref{clusters.fig}
lists all the possible scenarios leading to n-clusters in the steady state.  Ultimately, a cluster of occupied sites needs be bounded by
empty clusters on the left and right.  A series of alternating sites in between does not lead to an occupied cluster (case~(c) of Section~II) and at the least a two-site cluster ($\p\p$) needs to be present (Fig.~\ref{clusters.fig}a).   For the final cluster to attain length $n$, the combined lengths of the alternating sequences abutting the $\p\p$ seed should be $2(n-2)$. Because the seed could be initially in any of $n-1$ locations (including right next to the \qq\ blocks), the total contribution of this
case is
\[
P_a(n)=q^4p^2(n-1)(pq)^{n-2}\,,\qquad n\geq2\,.
\]

The initial seed could be three sites long (Fig.~\ref{clusters.fig}b), leading to a contribution
\[
P_b(n)=q^4p^3(n-2)(pq)^{n-3}\,,\qquad n\geq3\,.
\]
All other cases are exemplified by Fig.~\ref{clusters.fig}c.  The shaded block near the center denotes an arbitrary
sequence of sites with the property that it contains no two consecutive empty sites ($\q\q$).  All of the sites in such
a sequence would end up being occupied in the steady state.  Denoting the probability of an $m$-sequence
of this type by $\psi_m$, we can express the contribution of this last case as
\[
P_c(n)=q^4p^4\sum_{k=0}^{n-4}(k+1)(pq)^k\psi_{n-4-k}\,,\qquad n\geq4\,.
\]

To compute $\psi_n$, denote by $\xi_n$ the probability of an $n$-sequence that contains no $\q\q$'s and ends with $\p$,
and let $\eta_n$ denote similar sequences, but ending with $\q$.  The two quantities satisfy the recursion equations:
\[
\xi_{n+1}=(\xi_n+\eta_n)p;\quad\eta_{n+1}=\xi_nq\,.
\]
Using $\psi_n=\xi_n+\eta_n$, this can be recast into
\begin{equation}
\psi_{n+1}=p\psi_n+pq\psi_{n-1}\,.
\end{equation}
The solution is,
\begin{equation}
\psi_n=c_{+}r_{+}^n+c_{-}r_{-}^n;\quad r_\pm=\frac{p}{2}\pm\frac{\sqrt{p^2+4pq}}{2}\,,
\end{equation}
where $c_\pm$ are determined from the boundary conditions, $\psi_0=\psi_1=1$.

\begin{figure}[t]
\includegraphics[width=0.45\textwidth]{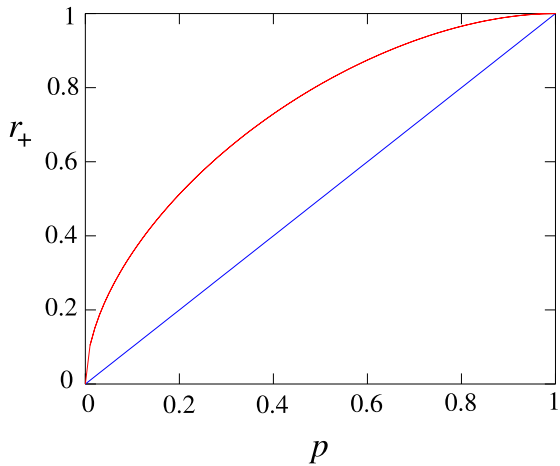}
\caption{(Color online) Probability of large occupied clusters:
The probability of $n$-clusters of occupied sites scales as $r_+^n$, in the steady state, as opposed to $\sim p^n$, of the
initial configuration.  Plotted are $r_+$ (curve) and $p$ (straight line) as a function of the initial density $p$.}
\label{rplus.fig}
\end{figure}

Although it is now straightforward to compute $P(n)=P_a(n)+P_b(n)+P_c(n)$, the resulting expression
is cumbersome and not particularly instructive.    Instead, we observe that, for large $n$, $P(n)\sim r_{+}^n$.
Since $r_{+}>p$ (Fig.~\ref{rplus.fig}), the probability for large clusters in the steady state is exponentially larger
than in the initial state.  For example, for small $p$, $r_{+}\sim\sqrt{p}$, and $P(n)\sim p^{n/2}$, compared to $\sim p^n$
of the initial state.

Despite the cumbersome form of $P(n)$, the average cluster size in the steady state yields a surprisingly simple expression:
\begin{equation}
\av{n_\p}_s=\frac{\sum nP(n)}{\sum P(n)}=\frac{2-p}{q^2}=(2-p)\av{n_\p}_0^2\,,
\end{equation}
where for the last equality we are using $\av{n_\p}_0=1/q$, for the sake of comparison with the initial state.

As a simple test of our results one can compute $p_s=\av{n_\p}/(\av{n_\p}+\av{n_\q})$, where the average length of clusters of empty sites, $\av{n_\q}=(2-q)/p^2$, is obtained from $\av{n_\p}$ by exchanging $p\leftrightarrow q$.  It is easy to confirm that this agrees with $p_s$ of Eq.~(\ref{ps}).  Also, $1/(\av{n_\p}+\av{n_\q})=\kappa_s$ of Eq.~(\ref{ks}), suggesting that there is no correlation between the lengths of adjacent occupied and empty clusters.  This seems remarkable, when contrasted with the correlations observed in $P(n)$ for large $n$.

\section{Dynamics}
To analyze the  transient behavior of the one-dimensional NCO model, consider the probability  $\pi_t$ that an initially occupied site
flips exactly $t$ times before settling down for good. $\pi_0=2p^2-p^3$ is simply the probability that the initially occupied site is stable,
i.e., it is the central site in one of the configurations: $\p\p\p,\,\q\p\p\,,\p\p\q$.  To $\pi_1$ contribute the $\p$-sites right next to the blocks \qq\  in cases (a), (b), and (c) of Section~II, resulting in
\[
\pi_1=2p^2q^2\frac{pq}{1-pq}+q^4p\frac{1+pq}{1-pq}\,.
\]
To $\pi_2$ contribute the $\p$'s two sites away from the blocks \pp\ in cases (a), (b) and (d), etc.  The general result is
\begin{equation}
\pi_t=2p^2q^2\frac{(pq)^t}{1-pq}+\omega(pq)^{t-1}\frac{1+pq}{1-pq}\,,
\end{equation}
where $\omega=p^4q$ or $q^4p$ for even and odd $n$,  respectively.
The analogous probabilities, $\sigma_t$, for a $\q$-site to flip exactly $t$ times before stopping are obtained, as usual,
by interchanging $p\leftrightarrow q$ in $\pi_t$.

Armed with $\pi_t$ and $\sigma_t$ one can compute various quantities of interest.  For example, the steady-state
density is
\[
p_s=\sum_{\tau=0}^{\infty}(\pi_{2\tau}+\sigma_{2\tau+1})\,,
\]
and it is easy to confirm the agreement of this formula with~(\ref{ps}).
More importantly, the probability that a site, of any kind, is still active at time $t$ (persistence), is
\begin{equation}
S_t=\sum_{\tau=t}^{\infty}(\pi_\tau+\sigma_\tau)\sim(pq)^t\,.
\end{equation}
The rapid exponential decay of $S_t$ is typical of other dynamical quantities.  In practical terms, the system settles almost instantly to the steady state.  An example of this effect can be seen in Fig.~\ref{ps.fig}, where the density of occupied sites
after merely one step, $p_1$, is compared to the density $p_s$ of the steady state.  An exception is the probability $P_t(n)$ for finding  $n$-clusters of occupied sites after $t$ steps. For  small $p$ (and large $n$), it  converges slowly to the steady state, as $P_t(n)\sim p^{\alpha_tn}$, with $\alpha_t=(t+1)/(2t+1)$.

\section{NCO on the Bethe lattice}
We now consider the NCO model on the Bethe lattice with coordination number $z\geq3$ (the line can actually be viewed
as a Bethe lattice with $z=2$).  The Bethe lattice with $z\geq3$ is far richer than the line, in that it sustains phase transitions:
being infinite-dimensional, its phase transitions are mean-field in character, as expected above the critical dimension.  For percolation, for example, a phase transition occurs at $p=p_c=1/(z-1)$. For $p>p_c$ there exist an {\em infinite} number of infinitely large connected clusters of occupied sites, which are characterized by mean-field critical exponents.  This is known as the {\em critical phase}. In regular lattices there is no critical phase, and instead  there is only {\em one} infinite cluster at $p>p_c$, known as the {\em percolative phase}~\cite{perc}.

For $z=3$ the neighborhood of each node consists of 4 sites (including the site itself) and according to the NCO rules a site retains its own state in the event of a tie.  It follows that, just as for the line, a two-site cluster is forever stable.  At $p>1/2$
the initial configuration possesses infinitely many infinite percolation clusters.   Those clusters will remain stable indefinitely,
in the NCO model, perhaps accreting even more of the neighboring sites as the updates take place.  Indeed, the threshold $p_c$ for the emergence of infinite clusters, in the NCO model,  is actually smaller than $1/2$ (of regular percolation).

We consider a ``root" node at $\ell=0$ and define layer $\ell$ as the set of nodes $\ell$ links away from the root.   Let $N_\ell$ denote the number of paths, from the root to layer $\ell$, that never encounter an initially stable empty site, that is, none of the empty sites on a path have an empty neighbor (either on, or branching away from the path).  All of the nodes on the
$N_\ell$ paths end up occupied in the steady state.
Following a similar reasoning to the one we applied for $\psi_n$ on the line, we find
\[
N_{\ell+1}=2pN_\ell+4p^2qN_{\ell-1}\,,
\]
leading to  $N_\ell\sim r^\ell$, with $r=p(1+\sqrt{1+4q})$.  The critical point for 
the emergence of infinite clusters in the NCO occurs therefore for $p_c\approx0.344446$ --- the root of $r=1$.  

To sum up, the NCO model on the $z=3$ Bethe lattice is very much like percolation in that there is a critical phase
of infinitely many infinite clusters at $p>p_c$, and only finite clusters at $p<p_c$.  That there is no transition to a percolative phase is clear from the fact that stable $\q\q$-clusters are seeded, initially, for all $p$ (other
than at the extremum, $p=1$), preventing the coalescence of all the infinite clusters into a single one.

The situation is even more interesting for $z=4$.  In this case, the neighborhood consists of 5 sites and a minimum of 2 occupied neighbors are required to stabilize a $\p$-site.  It is easy to see that under this condition the $z=4$ Bethe lattice does not support {\em finite} stable clusters.  However, the {\em infinite}
percolation clusters are stable also in the NCO model, so it follows that the existence of infinitely many infinite clusters of
occupied sites is guaranteed for $p>1/(z-1)=1/3$, and may occur somewhat sooner, at $p=p_{c1}\leq1/3$.  At $p<p_{c1}$,
infinite clusters fail to form and the steady state is uniformly empty, since finite occupied clusters are unstable and vanish as the NCO updates evolve.  The lattice becomes uniformly occupied,
at the steady state, for $p>p_{c2}=1-p_{c1}$, because of
the symmetry between empty and occupied sites under $p\leftrightarrow q$.  In other words, the NCO model has now {\em three} phases: no infinite clusters,
for $p<p_{c1}$; infinitely many infinite clusters, for $p_{c1}<p<p_{c2}$; and a single infinite cluster (consisting of the whole lattice), for $p>p_{c2}$.  Unlike regular percolation, there are no finite clusters in the non-critical phases.   The situation is similar for $z>4$.

\section{discussion}
In conclusion, we have presented an exact solution of the NCO model on the line.  This analysis is particularly interesting in view of the suspected connection between the NCO model and TIP~\cite{shao}.  It should be noted that while invasion percolation on the line is trivially defined and solved, it is not at all  clear how to implement the trapping condition --- that is,
superficially, there seems to be no difference between invasion percolation and TIP on the line, and neither displays the richness of results found for the NCO model in this paper.

The situation is even more interesting for Bethe lattices.  Just as in one dimension, there is no obvious way to distinguish between invasion percolation and TIP.  For $z=3$, the two-phase diagram of the NCO model is the same as that of {\em regular}
percolation (for any $z$).  The three phases found for the NCO model on the Bethe lattice with $z\geq4$ can be found in regular percolation, but only on hyperbolic  lattices and other nonamenable one-ended graphs~\cite{nags}.   Solving the NCO model on Bethe lattices is an appealing open problem. It remains ultimately unclear, however, how to connect the rich behavior of the NCO model on the line and on Bethe lattices to TIP.

\acknowledgements

I thank Erik Bollt, Chris Jizheng, and Wen Luosheng for many useful discussions and for encouraging this line of inquiry.


\begin{thebibliography}{9}
\bibitem{remark1} In~\cite{shao} it was argued that changing the node's state with
probability $p$ ($<1$), in the event of a tie, yields similar qualitative results to $p=0$ (our case).
\end{thebibliography}
\end{document}